# Short range order and topology of binary Ge-S glasses

I. Pethes[1,*], P. Jóvári[1], S. Michalik[2], T. Wagner[3,4], V. Prokop[3], I. Kaban[5], D. Száraz[1], A. Hannon[6], M. Krbal[3]

[1]Wigner Research Centre for Physics, Institute for Solid State Physics and Optics, Konkoly Thege út 29-33., H-1121 Budapest, Hungary

[2]Diamond Light Source Ltd., Harwell Science and Innovation Campus, Didcot, Oxfordshire, OX11 0DE, UK

[3]Center of Materials and Nanotechnologies (CEMNAT), Faculty of Chemical Technology, University of Pardubice, Legions Square 565, 530 02 Pardubice, Czech Republic

[4]Department of General and Inorganic Chemistry, Faculty of Chemical Technology, University of Pardubice, Studentska 573, 532 10 Pardubice, Czech Republic

[5]IFW Dresden, Institute for Complex Materials, Helmholtzstr. 20, 01069 Dresden, Germany

[6]ISIS Facility, Rutherford Appleton Laboratory, Didcot OX11 0QX, United Kingdom

Abstract

Short range order and topology of $Ge_xS_{100-x}$ glasses over a broad composition range ($20 \leq x \leq 42$ in at%) was investigated by neutron diffraction, X-ray diffraction, and Ge K-edge extended X-ray absorption fine structure (EXAFS) measurements. The experimental data sets were fitted simultaneously in the framework of the reverse Monte Carlo simulation method. It was found that both constituents (Ge and S) satisfy the Mott-rule in all investigated glasses: Ge and S atoms have 4 and 2 neighbours, respectively. The structure of these glasses can be described with the chemically ordered network model: Ge-S bonds are preferred; S-S bonds are present only in S-rich glasses. Dedicated simulations showed that Ge-Ge bonds are necessary in Ge-rich glasses. Connections between Ge atoms (such as edge-sharing $GeS_{4/2}$ tetrahedra) in stoichiometric and S-rich glasses were analysed. The frequency of primitive rings was also calculated.

1. Introduction

---

[*] Corresponding author. E-mail address: pethes.ildiko@wigner.hu

Ge-S glasses have been studied for several decades. In their paper in 1971 Kawamoto and Tsuchihashi [1] proposed that in S-rich $Ge_xS_{100-x}$ glasses ($x \leq 33.3$) 'the structure is based on a three-dimensional inorganic polymer of polymeric S chains cross-linked with Ge' while for glasses with $40 \leq x \leq 45$ Ge atoms can be found either in tetrahedral ($GeS_{4/2}$) or a rocksalt-type octahedral ($GeS_{6/6}$) environment where both Ge and S are sixfold coordinated. In this model there are no Ge-Ge bonds; sulfur atoms violate instead the 8-$N$ rule of Mott [2] to satisfy the valence requirements of Ge. From their XRD measurements on glasses with $x = 33.3$ and 42 Rowland et al. [3] deduced that 'it is extremely unlikely that $GeS_{6/6}$ octahedra are formed' in the $Ge_{42}S_{58}$ glass. In their study based on infrared and Raman spectroscopy, Lucovsky et al. concluded that short range order in $Ge_xS_{100-x}$ glasses ($10 \leq x \leq 45$) does satisfy the 8-$N$ rule [4]: Ge atoms have 4 nearest neighbours and S atoms are twofold coordinated. The authors found that the structure of the stoichiometric composition ($GeS_2$) can be described by the chemically ordered network model (CONM), in which the S atoms are tetrahedrally arranged around Ge atoms, and neither Ge-Ge nor S-S bonds are present. In sulfur-rich alloys, S-S bonds also appear, probably in the form of $S_8$ rings. They concluded that glasses in the Ge-rich region are also ordered chemically (only Ge-Ge and Ge-S bonds are formed).

Based on Raman and Mössbauer spectroscopy results Boolchand et al. [5] proposed a more complex model in 1986. They suggested that in $GeS_2$ ethane-like ($Ge_2(S_{1/2})_6$) clusters are also present beside the $Ge(S_{1/2})_4$ tetrahedra, thus the structure is 'qualitatively incompatible with chemically ordered 3D continuous random networks'. In Ge-rich glasses a third type of molecular clusters, the so-called double-layered or distorted rock-salt GeS microphase ($GeS_{6/6}$) can be present.

Since the above three studies several theoretical and experimental papers have been published for and against the chemically ordered network model. Raman spectroscopy was used in many of these investigations [6 – 14]. *Ab initio* molecular dynamics simulations using density functional theory were also applied to calculate the vibrational normal modes of cluster models and compute the associated Raman activities [10, 15]. In a more recent paper isotopically enriched ($^{72}Ge$, $^{74}Ge$, $^{76}Ge$) $GeS_{1.35}$ samples were studied with Raman spectroscopy and quantum-mechanical calculations using the hybrid B3LYP functional [13]. Raman studies agree on the origin of most observed Raman-active modes/frequencies (concerning e.g. the modes of a $Ge(S_{1/2})_4$ tetrahedron and two such tetrahedra connected by one or two common sulfur atoms (corner and edge-sharing tetrahedra, respectively)), but there is no consensus about the origin of some peaks, for example around 250 $cm^{-1}$, which is either assigned to the Ge-Ge pairs of ethane-like units [8, 13, 15] or double layered distorted rock-salt clusters [5, 12, 14] or both [7, 9] or threefold bonded S and Ge atoms in $SGe_3$-$S_{6/3}$ units [10].

EXAFS measurements using the K-edge of Ge support the fourfold coordination of Ge atoms in agreement with the Mott-rule [16 – 19]. Ge-S bond length was usually found to be around 2.20 – 2.23 Å. In Ge-rich alloys, the presence of Ge-Ge pairs is confirmed by EXAFS, with $r_{GeGe}$ = 2.48 – 2.52 Å, while no evidence of homonuclear bonding is found in stoichiometric $GeS_2$. The S K-edge EXAFS experiment of Armand et al. [20] (which is, as far as we know, the only published S K-edge EXAFS measurement of the Ge-S system) suggests S-S bonding in the S-rich samples, with a bond length around 2.05 – 2.06 Å, and a total coordination number of S around 2.

Ge-S glasses were also investigated by X-ray and neutron diffraction [3, 11, 14, 21 – 29]. From the measured intensities the structure factor can be calculated, whose Fourier transform gives the total pair correlation function. Unfortunately, from a single diffraction dataset the peaks of the total pair correlation function in most cases cannot be unambiguously assigned to the structural motifs. There is a consensus that the peak around 2.2 Å comes from the Ge-S pairs, in S-rich samples often together with the somewhat shorter S-S bonds [11, 23 – 25, 29]. However, the origin of the second peak in Ge-rich glasses around 2.44 Å is still an open question [14, 27]: it can either be assigned to Ge-Ge pairs from ethane-like units or according to the distorted rock-salt model, to long Ge-S bonds.

Molecular dynamics simulations are also used to calculate the pair correlation functions and structure factors of Ge-S glasses, most often the stoichiometric composition [30 – 33], but there are some studies on S-rich [34] and Ge-rich compositions as well [28, 35]. Unfortunately, the small simulation box size and the high cooling rates available in this type of simulation can affect the results, as was shown in Refs. [31, 32, 34]. The glass transition temperature and the number of homonuclear bonds decrease as the cooling rate applied in the simulation decreases [32]. Different simulation box sizes resulted in a different number of homonuclear bonds in Refs. [31, 34]. Moreover, the applied functional may also have a significant effect on the model structure.

The combination of different experimental methods and/or theoretical calculations can help to get a more reasonable description of the structure of these glasses. The reverse Monte Carlo (RMC) method [36] was applied to fit data from neutron diffraction measurement on $GeS_2$ in Ref. [29]. For $GeS_3$ glass diffraction and Ge K-edge EXAFS datasets were fitted in the framework of the RMC method [26], while a combination of DFT simulation with RMC refinement was presented in Ref. [28].

Besides the origin of the 250 cm$^{-1}$ frequency in the Raman spectra and the interpretation of the peak of the experimental pair correlation functions at 2.44 Å there are other open questions. One of them is the presence of homonuclear bonds in the stoichiometric $GeS_2$ glass. In the analogous $GeSe_2$ glass Se-Se and Ge-Ge homonuclear bonds were detected (see e.g. [37] or [38]). Concerning the $GeS_2$ glass, the results are not conclusive: homonuclear bonds were found in Refs. [5, 6, 8, 9, 15, 16, 31 – 33, 35], chemically ordered network, without the presence of Ge-Ge and/or S-S bonds was reported in Refs. [4, 7, 11, 17 – 20, 24, 25, 29, 30].

It is generally accepted that in glassy $GeS_2$ $GeS_{4/2}$ tetrahedra are connected partly by two common sulfur atoms (edge-sharing). The ratio of edge and corner-sharing tetrahedra is an open question, results scatter in the 0.16 – 0.47 range [9, 11, 25, 29, 30 – 32, 34].

The compositional dependence of the frequency of edge-sharing tetrahedra has also been investigated. It was found in the early EXAFS experiments [18] that the number of edge and corner-sharing $GeS_{4/2}$ units is the highest around the stoichiometric composition. The presence of edge-sharing tetrahedra in the S-rich compositions was suggested by Bychkov et al. [11]: they found that the fraction of these units decreases with increasing S content (from 0.44 in $GeS_2$ to 0.25 in $GeS_9$), but they exist even in $GeS_9$. Similar composition dependence was observed by high-resolution X-ray photoelectron spectroscopy (XPS): Golovchak et al. [39] reported edge-sharing tetrahedra in the $12.5 \leq x$ Ge content $Ge_xS_{100-x}$ glasses. The presence of edge-sharing tetrahedra in Ge-rich glasses was proposed by Bytchkov et al. [27] up to the highest Ge content (47 at.%) investigated by the authors.

In this study, we investigate the structure of four various $Ge_xS_{100-x}$ glasses over a broad composition range ($20 \leq x \leq 42$). For all compositions investigated two or three diffraction or EXAFS datasets were used to generate large scale structural models within the framework of RMC simulation. The resulting atomic configurations are used to calculate coordination numbers, and bond lengths along with other structural parameters (e.g. the ratio of edge-sharing tetrahedra). The uncertainty of structural parameters is also estimated and the results are compared with models proposed in the literature.

2. Experimental

A total of 10 g of elements with 5N purity was placed into a quartz ampoule, evacuated to $10^{-3}$ Pa, and subsequently sealed. Next, the ampoule was placed in a rocking furnace heated to 970 °C with a heating rate of 1 °C/min and then was held rocking for 36 h at this temperature. In a further step, the temperature was reduced to 700 °C with a cooling rate of 2 °C/min, then stop rocking for 1 h and subsequently the ampoule was quenched in cold water. Finally, samples in ampoules were annealed for 3 h at about 20 °C below the glass-forming temperature to release the internal stress.

The neutron diffraction measurement of $Ge_{20}S_{80}$ and $Ge_{33}S_{67}$ was carried out at the GEM diffractometer of Rutherford Appleton Laboratory (UK). Powdered samples were filled into thin-walled cylindrical vanadium cans (inner diameter 8.3 mm, wall thickness 0.04 mm). The height and width of the incident beam were 40 mm and 15 mm, respectively. Raw data were corrected for empty instrument background, scattering from the sample holder, detector dead time, sample absorption, and incoherent scattering. Scattering from a rod made of $V_{0.949}Nb_{0.051}$ null alloy was also measured for the purpose of normalisation.

High energy X-ray diffraction data of $Ge_{20}S_{80}$ and $Ge_{33}S_{67}$ were collected on the Joint Engineering, Environmental and Processing (I12-JEEP) beamline [40] at Diamond Light Source Ltd., the United Kingdom. The sample material was firstly ground and then loaded into a thin-walled borosilicate capillary of 1.5 mm in diameter. The capillary was illuminated by an X-ray beam of the energy of 100.046 keV and the size of $0.5 \times 0.5$ mm$^2$ for 300 seconds. The diffracted X-rays were detected by a flat-type Pilatus 2M CdTe detector positioned at a distance of 236.6 mm from the sample in transmission geometry. The energy and geometry calibration [41] together with the azimuthal integration of 2D diffraction data into the reciprocal space were performed using the DAWN software [42]. Raw intensity 1D curves were then corrected for background scattering (empty capillary and air contributions), sample absorption, fluorescence, and Compton scattering using standard procedures [43] to get only elastically scattered intensities from a sample. Finally, the intensity curves were normalised applying the Faber-Ziman formalism [44] to extract structure factors.

In addition to $Ge_{20}S_{80}$ and $Ge_{33}S_{67}$, $Ge_{25}S_{75}$ and $Ge_{42}S_{58}$ were also modelled in the present study using previously published data. Details of sample preparation and data collection can be found in Refs. [26] and [28].

3. Reverse Monte Carlo simulation

The reverse Monte Carlo simulation technique [36] is a robust modelling method to generate atomic configurations compatible with experimental data. Mostly neutron and X-ray diffraction structure factors are used as input but EXAFS, electron diffraction, or anomalous X-ray scattering data can also be fitted. Along with experimental information physical and chemical knowledge is also built into the models generated by RMC. Density, minimum interatomic distances, coordination numbers, and bond angle distributions can all be used to make the RMC-generated configurations more reliable. The RMC++ code [45] (version 2.3) was used in the present study. The EXAFS backscattering coefficients were calculated by the FEFF8.4 program [46].

*Unconstrained simulations* were carried out with boxes containing 20000 atoms. Densities were taken from [7]. The values used in the present work are listed in Table 1. Initially, the atoms were placed into the simulation box randomly, then they were moved around to satisfy the minimum interatomic distances. Cut-offs (nearest interatomic distances) were 2.3 Å, 2.04 Å, and 1.96 Å for bonding Ge-Ge, Ge-S, and S-S pairs, respectively. If Ge-Ge or S-S bonds were forbidden then the corresponding cut-offs were raised to 2.84 Å and 3.16 Å, respectively. The real space grid size was 0.08 Å. As Ge-S and S-S bond lengths are quite close to each other the first peaks of the corresponding partial pair correlation functions cannot be separated by fitting diffraction data with the reverse Monte Carlo method (at least the neutron diffraction data of the present study up to ~ 35 Å$^{-1}$ do not make it possible in case of $Ge_{20}S_{80}$). This resulted in artificial split S-S first peaks with a maximum around 2.06 Å and a second peak close to the Ge-S distance. Therefore a 'gap' was defined in models of S-rich glasses by setting $g_{SS}(r) = 0$ between 2.2 Å and 3.1 Å. Apart from the S-S gap unconstrained models rely only on diffraction data and basic physical information (density, minimum interatomic distances).

The uncertainties of the coordination numbers of unconstrained models were estimated by dedicated simulation runs in which average coordination constraints were used to deviate coordination numbers from the experimental values in steps of 5-10 %. Coordination numbers and nearest neighbour distances obtained by unconstrained simulation runs are given in Tables 2 and 3.

In the *constrained simulation runs* each Ge atom was forced to have 4 neighbours while S atoms had to have 2 neighbours. These constraints were usually satisfied by about 90-95 % of the atoms.

4. Results and discussion

   4.1. Fitting of the experimental data

Several different models were tested to get the most reliable models of the glasses. It was found that all data sets can be properly fitted using the chemically ordered network model. According to the CONM, in the $Ge_{33}S_{67}$ glass only Ge-S bonds were allowed, applying higher cut-off values for the Ge-Ge and S-S pairs. In the S-rich $Ge_{20}S_{80}$ and $Ge_{25}S_{75}$ glasses, the S-S pairs were also allowed beside the Ge-S pairs, but Ge-Ge bonding was forbidden. In the case of the $Ge_{42}S_{58}$ glass according to the CONM Ge-Ge and Ge-S bonds were allowed. The experimental ND and XRD data, together with the fits obtained by RMC simulations using the CONM model are shown in Figures 1 and 2.

   4.2. The structure of $Ge_{33}S_{67}$

This composition was investigated by unconstrained simulations. As expected, the Ge-Ge and S-S cut-offs can be raised to the non-bonding values (2.84 Å and 3.16 Å) without deteriorating the fit quality. The partial pair correlation functions obtained by the unconstrained simulation are shown in Figure 3. The first peak of $g_{GeGe}(r)$ is at about 2.91 ± 0.03 Å while the corresponding coordination number is 0.48 ± 0.2. This peak is well known in the literature and is attributed to the presence of edge-sharing $GeS_{4/2}$ tetrahedra both in glassy and crystalline $GeS_2$ [29, 47]. The main peak of $g_{GeGe}(r)$ is located at 3.53 ± 0.03 Å and the total (second neighbour) Ge-Ge coordination number calculated from 2.8 Å to 4.0 Å is 3.58. (It is to be noted that in $GeS_2$ the total second neighbour Ge-Ge coordination number is expected to be $4 - N_{ES}$ where $N_{ES}$ is the average number of edge-sharing $GeS_{4/2}$ tetrahedra around a Ge atom (0.48 ± 0.2).) The mean nearest neighbour Ge-S distance is 2.23 ± 0.01 Å and the coordination number calculated up to 2.56 Å Ge-S separation is 3.90 ± 0.1. The position of the first S-S peak is around 3.3 Å. This value is clearly below the mean S-S distance in $GeS_{4/2}$ tetrahedra (3.63 Å). This is partly due to the distortion of the $GeS_{4/2}$ tetrahedra. However, topologically distant (non-second neighbour) S-S pairs also contribute to the first peak of $g_{SS}(r)$ (see Figure 4), similarly to crystalline $GeS_2$ where S-S distances within and between $GeS_{4/2}$ tetrahedra overlap.

### 4.3. The structure of $Ge_{20}S_{80}$ and $Ge_{25}S_{75}$

Partial pair correlation functions of $Ge_{20}S_{80}$ and $Ge_{25}S_{75}$ are shown in Figure 5. The mean Ge-S distance is around 2.22 Å ± 0.02 Å in both compositions. This value agrees well with the Ge-S bond length found in previous studies using neutron or X-ray diffraction [29, 48] or EXAFS [18]. The S-S distance is 2.06 ± 0.02 Å, which is close to the S-S bond length found in amorphous and liquid S [49]. The main Ge-Ge peak position (3.40 ± 0.03 Å) agrees well with the distance of Ge atoms sharing a common S neighbour. The Ge-Ge (second neighbour) coordination number calculated up to 3.9 Å is 2.31 and 2.77 for $Ge_{20}S_{80}$ and $Ge_{25}S_{75}$, respectively. The Ge-Ge peak position along with the coordination number shows that corner-sharing of $GeS_{4/2}$ tetrahedra exists already in $Ge_{20}S_{80}$ where sharing of S atoms can be avoided. It should be noted that the Ge-Ge coordination number corresponding to the peak at ~ 2.90 Å is around 0.2 – 0.3. This value is around the sensitivity of our approach and the Ge-Ge cut-off can be raised to above 3 Å without the deterioration of fit quality. Thus, edge-sharing of $GeS_{4/2}$ tetrahedra in $Ge_{20}S_{80}$ and $Ge_{25}S_{75}$ cannot be verified by the present data.

### 4.4. The structure of S-poor $Ge_{42}S_{58}$

In the case of $Ge_{42}S_{58}$ composition Ge K-edge EXAFS data were also fitted beside the two diffraction data sets. Partial pair correlation functions are shown in Figure 6. It was found that the S-S cut-off can be increased to its non-bonding value (3.16 Å) without altering the fit quality. The first S-S peak is around 3.65 Å. This value corresponds to the distance of two S atoms binding to a common Ge neighbour. The Ge-S distance is about 2.22 ± 0.02 Å, as in the S-rich compositions. The first peak of $g_{GeGe}(r)$ is at 2.44 ± 0.02 Å. Similar Ge-Ge bond lengths were obtained by EXAFS of Ge-rich Ge-Se films (2.45 – 2.48 Å, [50]), or by fitting diffraction and EXAFS data of glassy $Ge_{23.6}Te_{76.4}$ (2.45 Å [51]).

In a recent paper by Sakaguchi et al. [14], the authors proposed a model for $Ge_{40}S_{60}$ glass in which two Ge-S bond lengths (2.22 and 2.44 Å) were assumed, while Ge-Ge bonds were not present. The authors fitted their X-ray diffraction data with this model. Previously several EXAFS [17 – 19] or diffraction studies [27] suggested the presence of Ge-Ge bonds.

We investigated the necessity of Ge-Ge bonds by using a high Ge-Ge cut-off value (2.84 Å). Similarly to [14], it was found that the XRD dataset alone can be fitted equally well with and without Ge-Ge bonds. However, neutron diffraction and Ge K-edge EXAFS data cannot be reproduced without Ge-Ge bonds (see Figure 7 for the EXAFS fit). These observations all evidence that the pair distances around 2.45 Å should be attributed to Ge-Ge bonds.

### 4.5. Medium range order and topology analysis

Two edge-sharing $GeS_{4/2}$ tetrahedra form a four-membered (Ge-S-Ge-S) ring. In general, clusters can be determined as a group of atoms connected through a chain of nearest neighbours. Chains whose first and last atoms are connected are called cycles. The size of a cycle is the shortest path visiting all atoms in the cycle. Primitive rings are defined as cycles that cannot be decomposed into smaller cycles (see [52]). The distribution of primitive rings in Ge-S glasses was investigated by analysing the obtained particle configurations by a purpose-written program based on the NetworkX Python package [53].

In this analysis, the particle configurations obtained in constrained simulation runs were studied (Ge and S atoms were forced to have 4 and 2 neighbours, respectively). 3-membered rings (which appear as an artefact of the RMC method) were eliminated by forbidding the ~ 60° bond angles. Reference (called 'random' hereafter) configurations were obtained by hard sphere (no data) simulations using the same coordination number and bond angle constraints. To estimate the uncertainty of results both 'random' runs and those fitting experimental data were repeated 6 times, starting from different initial configurations.

The size distribution of primitive rings in the Ge-S glasses is shown in Figure 8. The values obtained for the 'random' models are also shown for comparison. The frequencies of 5 and 6-membered rings are significantly higher than in the 'random' configurations. In the case of the stoichiometric composition, only alternating Ge-S-Ge-S (-Ge-S), shortly A-B type rings, are present (because only Ge-S bonds are allowed), thus only even membered rings can occur. The number of 6-membered rings is more than 2 times higher in the data-fitted configuration than in the 'random' model. In the S-rich compositions, the 4-membered rings are also more frequent than in the 'random' case.

In non-stoichiometric configurations, many other ring types can be found, in addition to alternating A-B type rings. Since in these glasses, Ge-Ge or S-S bonds are also present, the number of the Ge and S atoms in the ring can be different. Furthermore, a ring of the same size can be produced in several ways, e. g. a 6-membered ring may contain 1 Ge and 5 S atoms, 2 Ge and 4 S atoms, 3 Ge and 3 S atoms or 6 S atoms, etc. (Moreover, the sequence of the components can be different, e. g. Ge-S-S-Ge-S-S or Ge-S-Ge-S-S-S). The frequency of these rings was calculated in both the data-fitted model and the 'random' model, see Figure 9. It was found that the differences in the ring size distribution originate mostly from the alternating rings: in the case of the S-rich compositions, the 4 and 6-membered A-B type rings are significantly more frequent. However, in the Ge-rich $Ge_{42}S_{58}$ glass, the 4-membered Ge-S-Ge-S rings are less frequent in the data-fitted configurations than in the 'random' configurations. Among the 5-membered rings, the ring types closest to the alternating ring (Ge-S-Ge-S-S ring in S-rich glasses and S-Ge-S-Ge-Ge in the Ge-rich glass) are more frequent than in the 'random' model.

5. Conclusions

Short range order of binary $Ge_xS_{100-x}$ ($20 \leq x \leq 42$) was studied by fitting X-ray and neutron diffraction and EXAFS data in the framework of reverse Monte Carlo simulation method. It was revealed that the compositions investigated are chemically ordered (there are no S-S bonds in S-poor compositions while Ge-Ge bonds exist only in Ge-rich glasses) and satisfy the $8 - N$ rule within the experimental uncertainty. The coordination number corresponding to edge-sharing $GeS_{4/2}$ tetrahedra in $Ge_{33}S_{67}$ is $0.48 \pm 0.2$. It was shown that the experimental data sets of the Ge-rich composition cannot be fitted properly without the presence of Ge-Ge pairs. It was found that the 5 and 6-membered primitive rings are more frequent than they would be in a random model, which satisfies the requirements of the $8 - N$ rule and the chemically ordered network model. In S-rich Ge-S glasses alternating rings are the most preferred.

Acknowledgment

This work was carried out with the support of Diamond Light Source, instrument I12-JEEP (nt28087-1). I. P. and P. J. were supported by the ELKH (Eötvös Loránd Research Network) Grant No. SA-89/2021. M. K., V. P and T. W. thank the Czech Science Foundation (Grant 20-23392J) and the Ministry of Education, Youth and Sports (LM2018103).

**Table 1** Investigated compositions, their densities, and the fitted datasets

|  | Density [g/cm$^3$] | Number density [atoms/Å$^3$] | Fitted datasets |
|---|---|---|---|
| Ge$_{20}$S$_{80}$ | 2.5 | 0.0374 | ND, XRD |
| Ge$_{25}$S$_{75}$ | 2.65 | 0.0378 | ND, XRD |
| Ge$_{33}$S$_{67}$ | 2.7 | 0.0358 | ND, XRD |
| Ge$_{42}$S$_{58}$ | 3.2 | 0.0393 | ND, XRD, Ge EXAFS |

**Table 2** Coordination numbers of the investigated Ge-S glasses obtained by unconstrained simulation runs (The uncertainty of the coordination numbers was estimated by dedicated constrained simulation runs, see text.)

|  | $N_{GeGe}$ | $N_{GeS}$ | $N_{SGe}$ | $N_{SS}$ | $N_{Ge}$ | $N_S$ |
|---|---|---|---|---|---|---|
| Ge$_{20}$S$_{80}$ | - | 3.74 (3.5 – 4.5) | 0.94 (0.88 – 1.12) | 1.01 (0.6 – 1.3) | 3.74 (3.5 – 4.5) | 1.95 (1.65 – 2.15) |
| Ge$_{25}$S$_{75}$ | - | 3.78 (3.65 – 4.1) | 1.26 (1.2 – 1.36) | 0.9 (0.4 – 1.2) | 3.78 (3.65 – 4.1) | 2.16 (1.76 – 2.4) |
| Ge$_{33}$S$_{67}$ | - | 3.90 ± 0.1 | 1.95 ± 0.05 | - | 3.90 ± 0.1 | 1.95 ± 0.05 |
| Ge$_{42}$S$_{58}$ | 1.25 ± 0.25 | 2.54 ± 0.15 | 1.84 ± 0.1 | - | 3.79 ± 0.15 | 1.84 ± 0.1 |

**Table 3** Nearest neighbour distances of the investigated Ge-S glasses obtained by unconstrained simulation runs

|  | $r_{GeGe}$ | $r_{GeS}$ | $r_{SS}$ |
|---|---|---|---|
| Ge$_{20}$S$_{80}$ | - | 2.23 ± 0.02 Å | 2.04 ± 0.02 Å |
| Ge$_{25}$S$_{75}$ | - | 2.23 ± 0.02 Å | 2.06 ± 0.02 Å |
| Ge$_{33}$S$_{67}$ | - | 2.22 ± 0.01 Å | - |
| Ge$_{42}$S$_{58}$ | 2.45 ± 0.02 Å | 2.22 ± 0.02 Å | - |

Figures

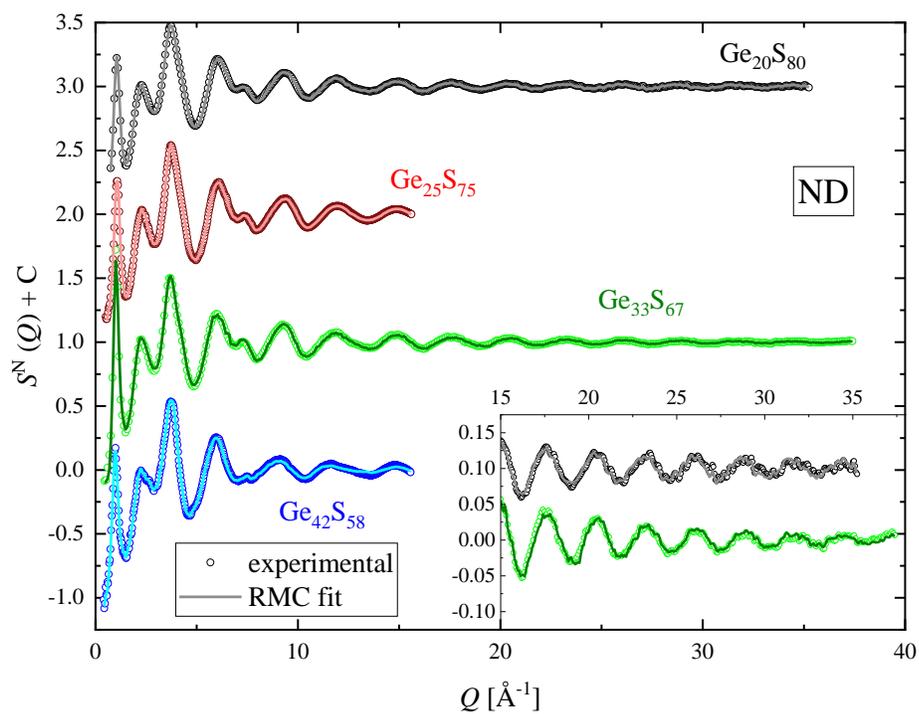

**Figure 1** Neutron diffraction structure factors (symbols) and fits (lines) of the Ge-S glasses. (The curves are shifted vertically for clarity.) The inset is an enlargement of the curves at high $Q$ values.

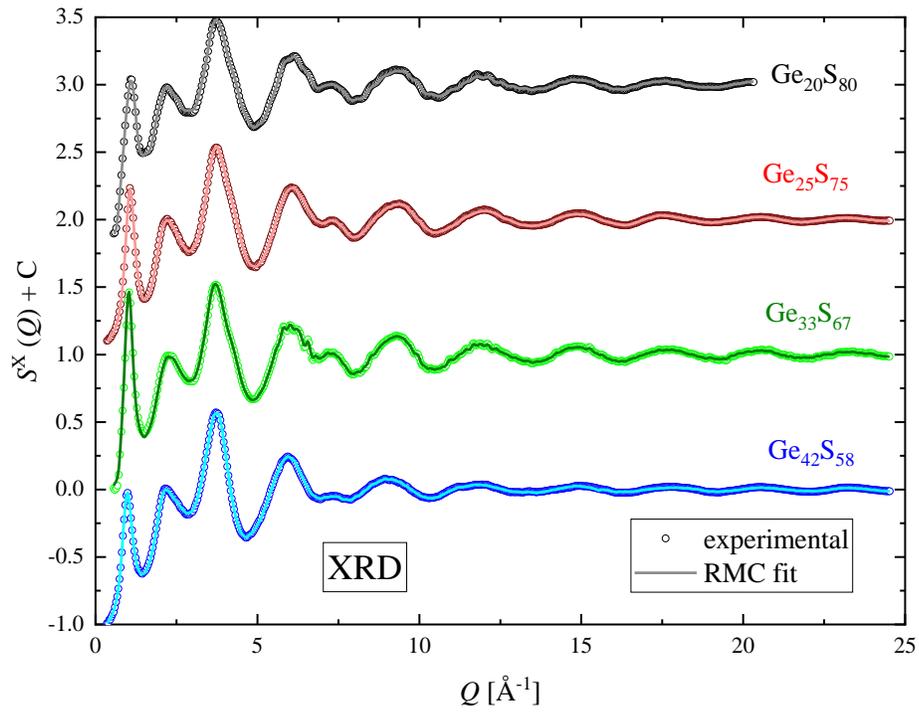

**Figure 2** X-ray diffraction structure factors (symbols) and fits (lines) of the Ge-S glasses. (The curves are shifted vertically for clarity.)

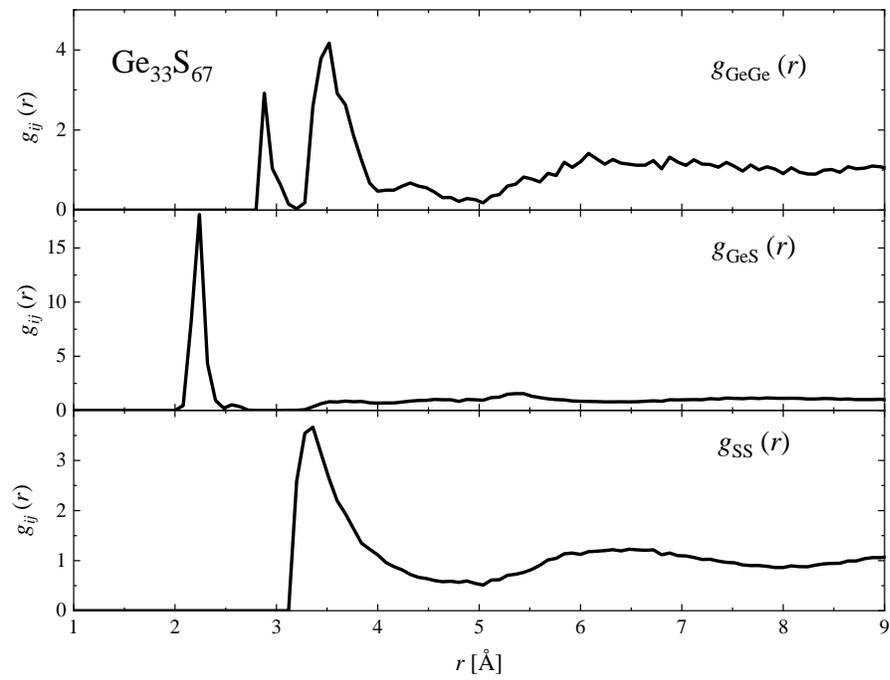

**Figure 3** Partial pair correlation functions of $Ge_{33}S_{67}$ glass.

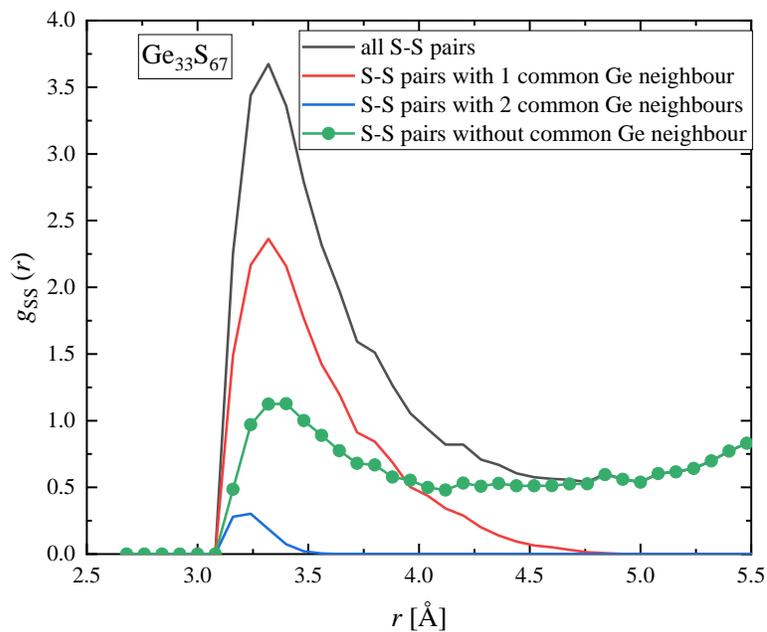

**Figure 4** Decomposition of the first peak of $g_{SS}(r)$ of $Ge_{33}S_{67}$. S-S pairs with 1 or 2 Ge neighbours in common form the edges of $GeS_{4/2}$ tetrahedra. S-S pairs without shared Ge neighbours are topologically distant pairs.

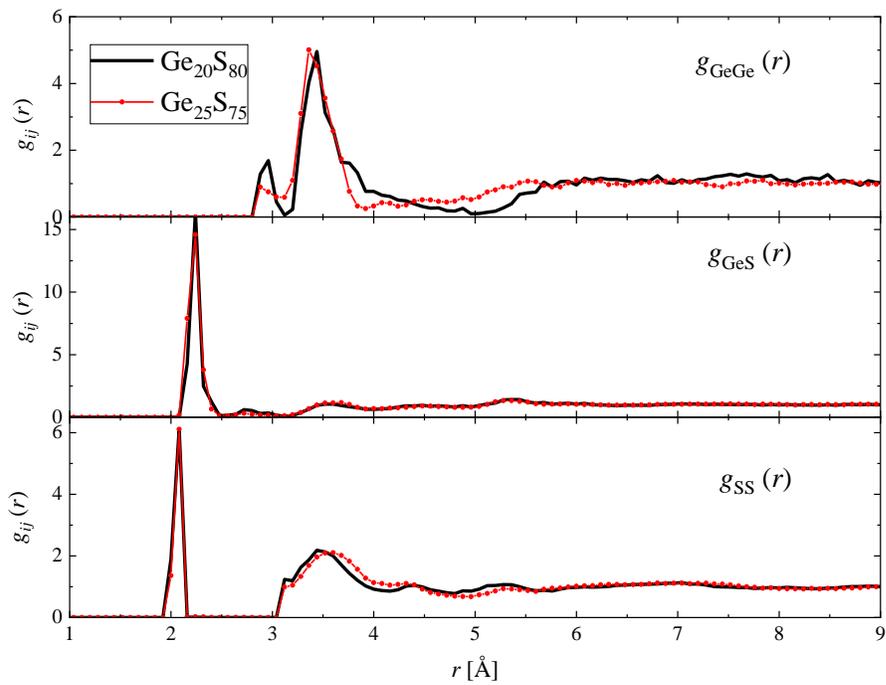

**Figure 5** Partial pair correlation functions of $Ge_{20}S_{80}$ and $Ge_{25}S_{75}$ glasses.

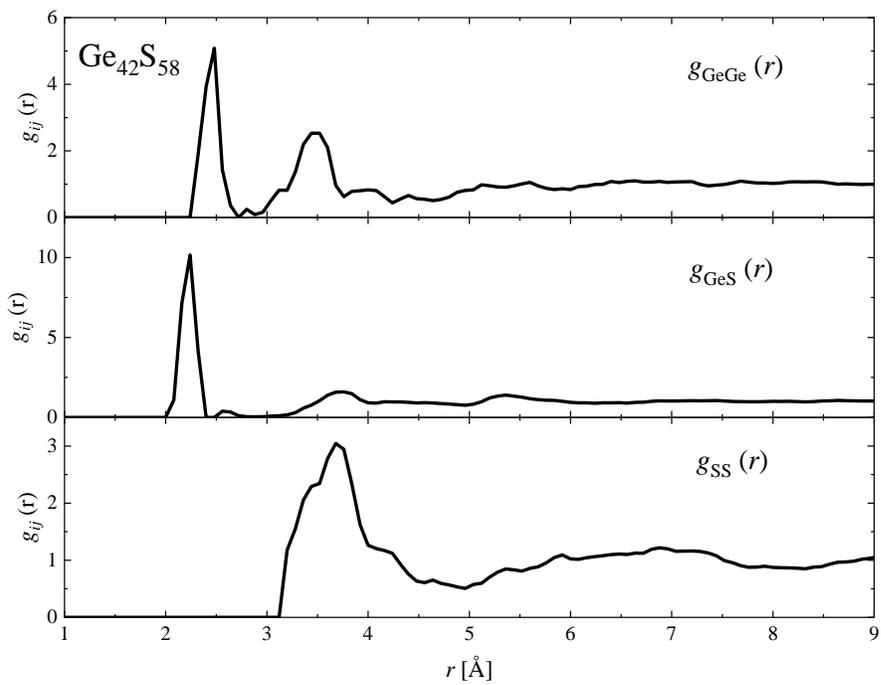

**Figure 6** Partial pair correlation functions of $Ge_{42}S_{58}$ glass.

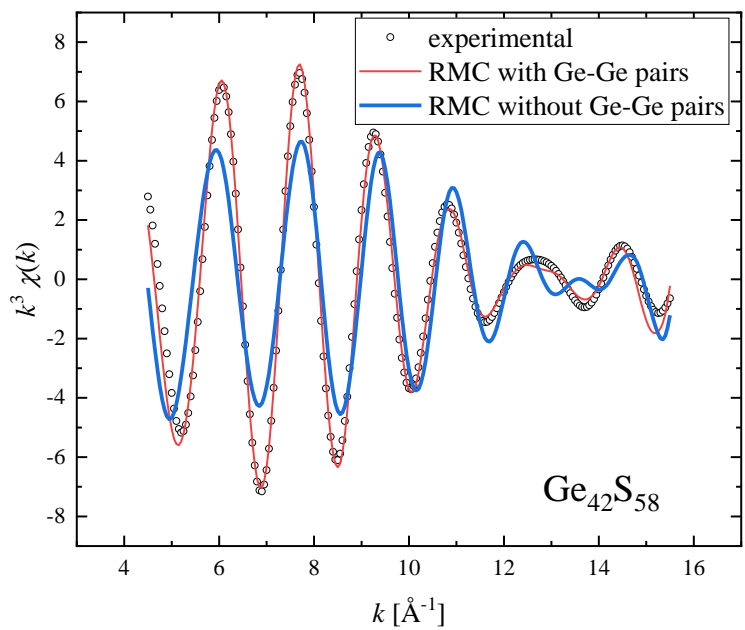

**Figure 7** Ge K-edge EXAFS fits of $Ge_{42}S_{58}$ with and without Ge-Ge bonds.

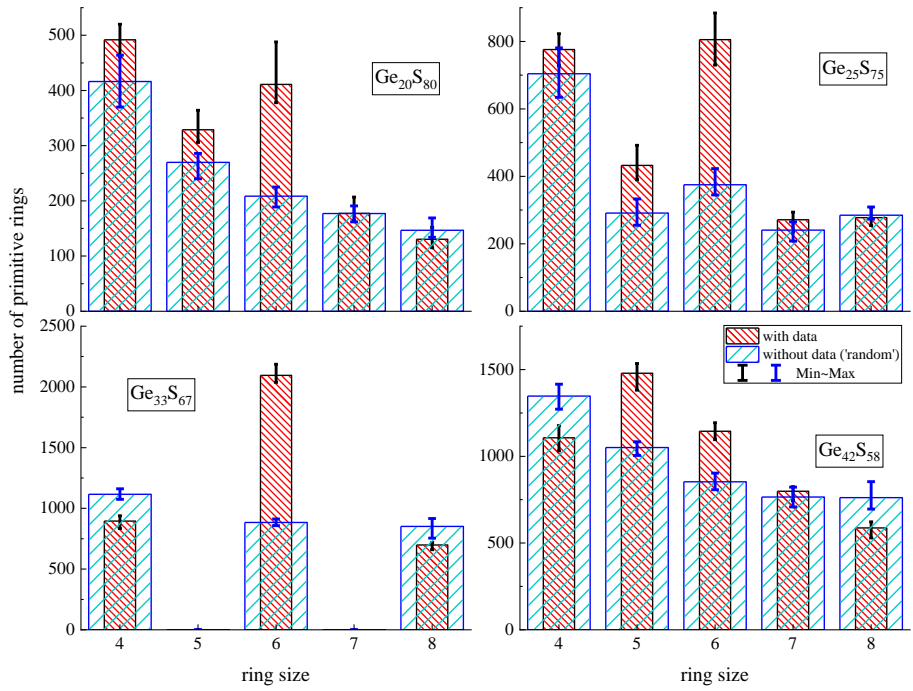

**Figure 8** Size distribution of primitive rings in different Ge-S glasses. The results of the 'random' model are also shown for comparison. Uncertainties are estimated by the difference of maximum and minimum values obtained in 6 different simulation runs.

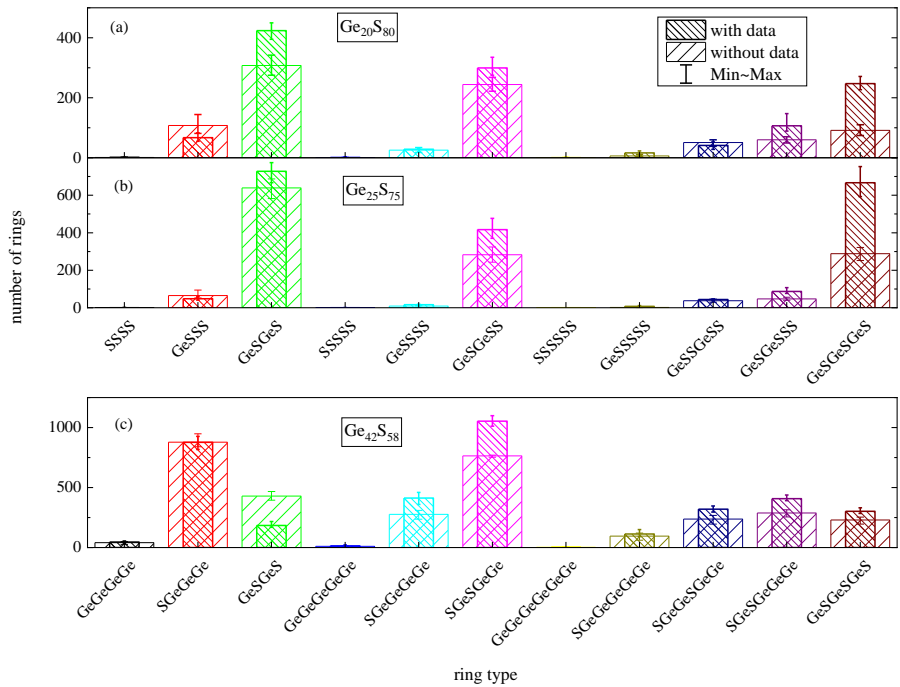

**Figure 9** Distribution of the different types of primitive rings in Ge-S glasses. (In $Ge_{33}S_{67}$ glass only A-B type rings are present, their numbers are in Fig. 8.) The results of the 'random' model are also shown for comparison. Uncertainties are estimated by the difference between the maximum and minimum values obtained in 6 different simulation runs. Note that the possible ring types are different in the S-rich compositions (upper two panels) and in the Ge rich glass (lower panel).